\documentclass[aps,prl,twocolumn,superscriptaddress,%showpacs %twocolumn
,floatfix]{revtex4}
\usepackage{graphicx}
\usepackage{amsmath,amssymb}

\begin{document}

\title{Hyper-entanglement of photons emitted by a quantum dot}

\author{Maximilian Prilm\"uller}
\affiliation{Institut f\"ur Experimentalphysik, Universit\"at Innsbruck, Technikerstra{\ss}e 25, 6020 Innsbruck, Austria}

\author{Tobias Huber}
\affiliation{Institut f\"ur Experimentalphysik, Universit\"at Innsbruck, Technikerstra{\ss}e 25, 6020 Innsbruck, Austria}

\author{Markus M\"uller}
\affiliation{Institut f\"ur Halbleiteroptik und Funktionelle Grenzfl\"achen and Center for Integrated Quantum Science and Technology (IQST) and SCoPE, Universit\"at Stuttgart, Allmandring 3, 70569 Stuttgart, Germany}

\author{Peter Michler}
\affiliation{Institut f\"ur Halbleiteroptik und Funktionelle Grenzfl\"achen and Center for Integrated Quantum Science and Technology (IQST) and SCoPE, Universit\"at Stuttgart, Allmandring 3, 70569 Stuttgart, Germany}

\author{Gregor Weihs}
\affiliation{Institut f\"ur Experimentalphysik, Universit\"at Innsbruck, Technikerstra{\ss}e 25, 6020 Innsbruck, Austria}

\author{Ana Predojevi\'{c}}
\email[]{ana.predojevic@uni-ulm.de}
\affiliation{Institut f\"ur Experimentalphysik, Universit\"at Innsbruck, Technikerstra{\ss}e 25, 6020 Innsbruck, Austria}
\affiliation{Institute for Quantum Optics, Albert-Einstein-Allee 11, University of Ulm, 89081 Ulm, Germany}

\begin{abstract}
Entanglement is a unique quantum mechanical attribute and a fundamental resource of quantum technologies. Entanglement can be achieved in various individual degrees of freedom, nonetheless some systems are able to create simultaneous entanglement in multiple degrees of freedom - hyper-entanglement. A hyper-entangled state of light represents a valuable tool capable of reducing the experimental requirements and resource overheads and it can improve the success rate of quantum information protocols. 
Here, we report on demonstration of polarization and time-bin hyper-entangled photons emitted from a single quantum dot. We achieved this result by applying resonant and coherent excitation on a quantum dot system with marginal  fine structure splitting. Our results yield fidelities to the maximally entangled state of 0.81(6) and 0.87(4) in polarization and time-bin, respectively. 
\end{abstract}

%\pacs{03.67.-a, 03.67.Bg, 42.50.Dv} %32.80.Qk

\maketitle

%\textbf{Entanglement is a unique quantum mechanical attribute and a fundamental resource of quantum technologies. Entanglement can be achieved in various individual degrees of freedom, nonetheless some systems are able to create simultaneous entanglement in multiple degrees of freedom - hyper-entanglement. A hyper-entangled state of light represents a valuable tool capable of reducing the experimental requirements and resource overheads and it can improve the success rate of quantum information protocols. 
%Here, we report on demonstration of polarization and time-bin hyper-entangled photons emitted from a single quantum dot. We achieved this result by applying resonant and coherent excitation on a quantum dot system with marginal  fine structure splitting. Our results yield fidelities to the maximally entangled state of 0.81(6) and 0.87(4) in polarization and time-bin, respectively.}

Quantum dots are semiconductor emitters of quantum light, which makes them material-wise compatible with today's information technologies. Furthermore, the latest advances in the design and implementation of quantum dots shows their competence to efficiently deliver indistingishable single photons \cite{Pascale,Pan,Unsleber} and photon pairs with high degree of entanglement \cite{Trotta, longtb}. These achievements combined with the possibility of photon storage \cite{Rare} show the potential of quantum dots to become building blocks of a quantum network \cite{Kimble}. Due to their discrete energy level structure quantum dots are inherently antibunched single photon \cite{gauss} sources with sub-Poissonian statistics \cite{turnstile} which allows them  to produce very pure single photon states \cite{Pascale,Pan}.

The application of entanglement of photons includes quantum communications \cite{Briegel, Ekert}, where it can be used as resource in information exchange protocols like teleportation \cite{bennett} and entanglement swapping  \cite{Zhukowski}. In addition, entanglement is an essential element of linear optical quantum computing \cite{klm}.  The entanglement-enhanced quantum communication schemes such as ultra-dense coding \cite{inn} and teleportation \cite{bennett} enable us, respectively, to transmit two bits in one qubit or securely communicate a quantum state. In such communication schemes the Bell-state-measurements are the crucial element. The simplest realization of a Bell-state measurement uses interference of two photons at a beam-splitter and has the disadvantage that it is efficiency limited \cite{Lutkenhaus, Vaidman}. The states of light that exhibit entanglement in more than one degree of freedom - hyper-entangled states \cite{Kwiat} can be used to perform a complete Bell measurement using linear optics %on photons that already posses quantum correlations
\cite{completeBell, slitty_eyes}. In addition, they are specifically valuable in lowering the resources overhead \cite{Graham} or for increasing the success rate \cite{Boyd} in the teleportation scheme.

Entanglement of photons emitted by quantum dots has been demonstrated in polarization \cite{Akopian, Young, Trotta, Emanuele, pol4, wires} and time-bin degrees of freedom \cite{time-bin}. The requirements for achieving a high degree of entanglement differ for the two approaches. High degree of polarization entanglement requires the absence of fine structure splitting of the quantum dot's exciton states. This is best achieved by post-growth modification and control of the quantum dot's energy levels \cite{Trotta} or by using alternative growth methods to self-assembly \cite{Emanuele}. In contrast, time-bin entanglement can be achieved on any quantum dot system even if the zero fine structure splitting condition is not fulfilled. Nonetheless, encoding in time-bin requires two-photon resonant excitation of the biexciton -- a method that allows for the coherent generation of exclusively photon pairs \cite{twophoton}.

\begin{figure*}[!ht]
\includegraphics[width=0.55\linewidth]{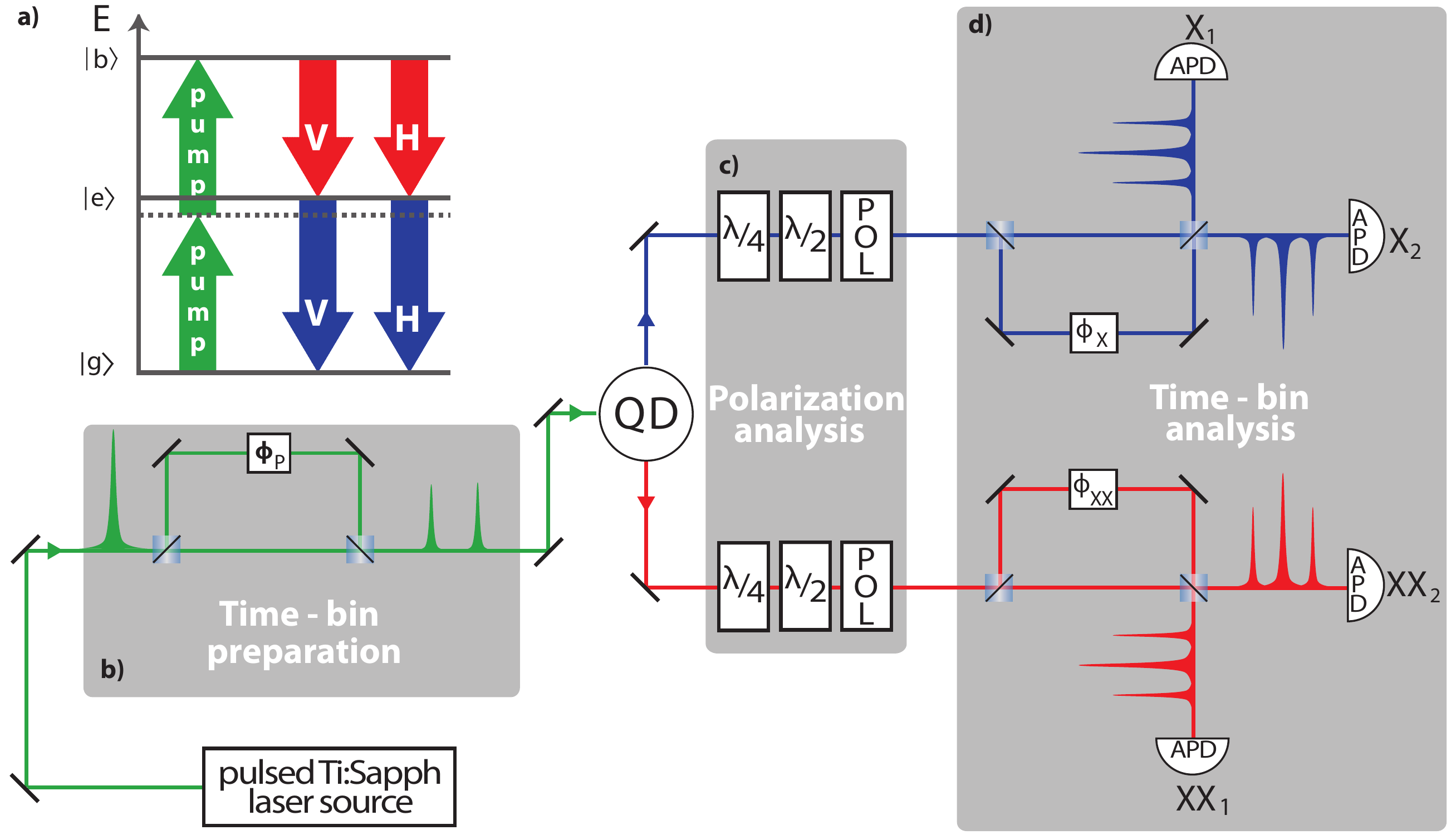}
\caption{(a) Energy scheme of the quantum dot. A laser coherently couples the ground, $|g\rangle$, to the biexciton state, $|b\rangle$, through two-photon resonant excitation. The excitation process is carried out via a virtual state (dashed line). The quantum dot system decays to the ground state via exciton state, $|b\rangle$, emitting the biexciton - exciton cascade. Due to vanishing fine structure splitting the two orthogonally polarized decay paths are indistinguishable and the emitted photons are entangled in polarization. b) Every excitation sequence consists of two phase-related excitation pulses that are required to generate entanglement in time bin. The phase between the pulses, $\phi_{p}$ is imposed by an unbalanced Michelson interferometer. The emission probability for each pulse is kept low to minimize the double excitations. c) The polarization analysis elements ($\lambda/4$ and $\lambda/2$ wave-plates and polarisers) were placed before the time-bin analysis. d) Both the generation and the analysis of the time-bin entanglement require unbalanced Michelson interferometers. The delay of the interferometer should be longer than any coherence in the quantum dot system, in our case 3~ns, which is 4.5 (10) times longer than the life time of the exciton (biexciton) photon. The entanglement is measured in the coincidence signal between a detector detecting biexciton photon, $\mathrm{XX}_{1,2}$, and a detector detecting an exciton one, $\mathrm{X}_{1,2}$.}
\label{2p}
\end{figure*}

Here, we report on the generation of hyper-entangled photon pairs emitted by a single quantum dot. The state that we created exhibits simultaneous entanglement in two degrees of freedom: polarization and time bin. To obtain it we used a quantum dot system that was excited resonantly by means of two-photon resonant excitation of the biexciton \cite{twophoton}, schematically depicted in Fig.~\ref{2p}a. In such an excitation process a quantum dot is resonantly driven from the ground to the biexciton state using a two-photon resonance at an energy equal to $(E_{xx}+E_{x})/2$, where $E_{xx}$ and $E_{x}$ are the energies of the biexciton and the exciton emission, respectively.  The excitation was carried out by a sequence of two pulses \cite{brendel}, so-called \textit{early} and \textit{late} pulse. The phase of the pump interferometer, $\phi_p$, (schematically depicted in Fig.~\ref{2p}b) that generated the excitation pulse sequence defined the phase between the early and the late pulse. Since the time-bin entanglement originates in interference of probability amplitudes for the system to be excited by the early or the late pulse \cite{franson}, the pump phase, $\phi_{p}$, directly affects the phase of the entangled state.

The pulse area of the excitation pulses was chosen such that we excite the quantum dot with low probability (6$\%$ or approximately a $\pi/15$-pulse). Due to the low excitation probability, the quantum dot was on average excited by only one of the two excitation pulses. The pulse length of the excitation pulse was chosen to be 20~ps in order to maximally suppress the single exciton probability amplitude while simultaneously giving maximal probability to coherently drive the ground-biexciton superposition \cite{longtb}. The single exciton excitation probability would  become dominant for shorter pulse lengths, due to the increased laser pulse spectral width. On the other hand, very long laser pulses would enable the double excitations of quantum dot within the same laser pulse, an event that would also reduce the coherence of excitation and with it the degree of entanglement.  Upon excitation to the biexciton state the quantum dot system decays to the ground state emitting a pair of photons. Due to this specific excitation method involving two low excitation probability pulses and the absence of fine structure splitting, the emitted pair of photons was entangled in two different degrees of freedom, i.e. hyper-entangled. Figure 1. shows four main elements of the experimental implementation: the two-photon resonant excitation level scheme, the generation of the excitation pulses, the polarization analysis (Fig.~\ref{2p}c) and the interferometers for analysis of time-bin entanglement (Fig.~\ref{2p}d).

\begin{figure*}[!ht]
\includegraphics[width=0.68\linewidth]{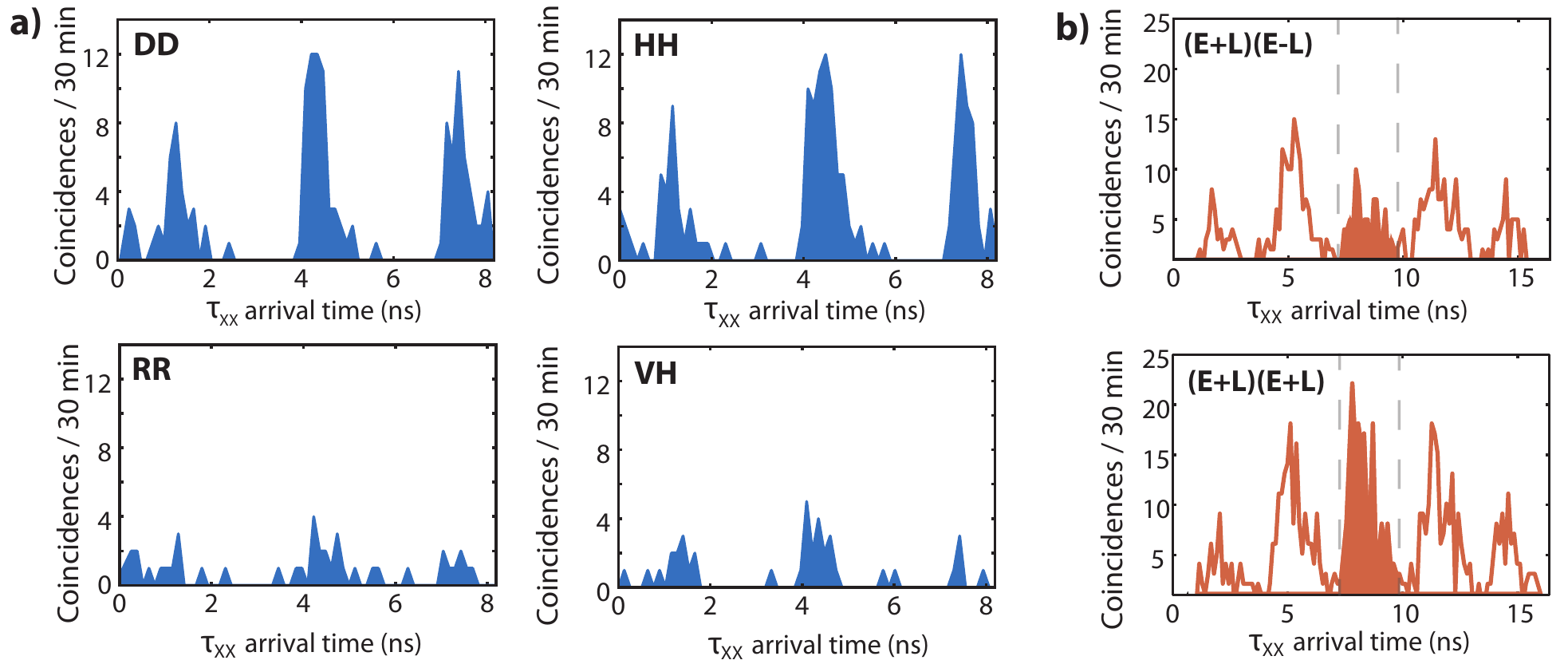}
\caption{(a) Projective measurements in polarization basis (DD, HH, RR, and VH). Each excitation pulse generates photons that after time-bin analysis interferometers can have three arrival times. The overall number of coincidence clicks in each projection we obtain by summing over all three arrival times. The $\tau_{xx}$ arrival is determined with respect to the excitation laser. (b) Projections in time-energy basis. This type of projection gives more complex result consisting of five peaks. If entanglement is present, the middle peak will change its hight as function of the relative phase between the analysis interferometers. The number of coincidence clicks is here determined by summing over the central peak (marked in orange and delimited by vertical dashed lines. Also here, the $\tau_{xx}$ arrival is determined with respect to the excitation laser. }
\label{peaks}
\end{figure*}

\begin{figure}[t]
\includegraphics[width=1\linewidth]{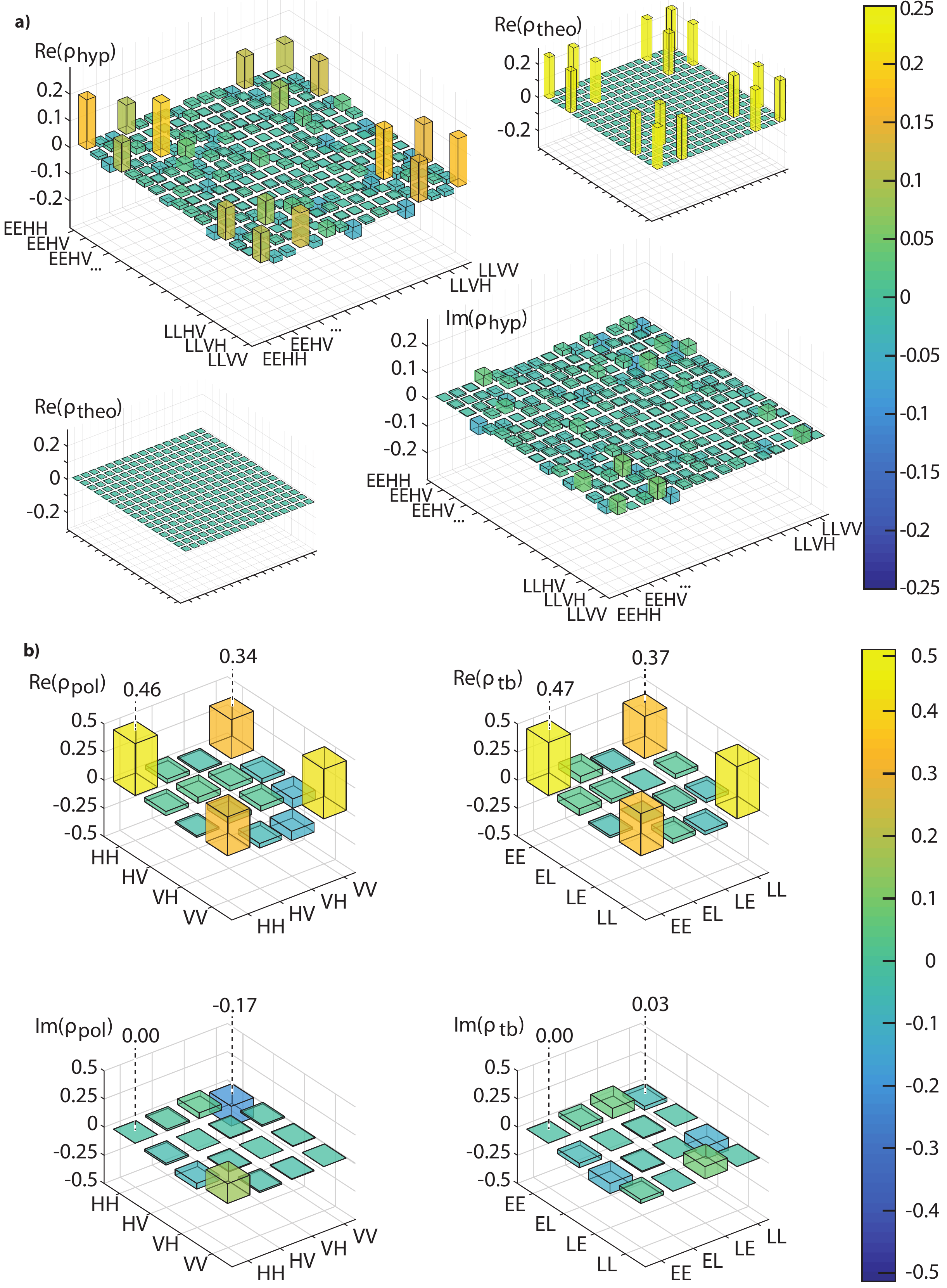}
\caption{(a) Real and imaginary part of the reconstructed density matrix for the polarization and time-bin hyper-entangled state. The insets up left and down right show, respectively, the real and the imaginary part of the density matrix for theoretically predicted state. (b) Real and imaginary part of the reconstructed density matrices for entanglement in polarization (left) and time bin (right). Here H and V stand for the basis states in the polarization entanglement measurement while E and L are the basis states of the time-bin measurement. The theoretically expected states in these measurements were $\frac{1}{\sqrt2}(|H\rangle_{XX}|H\rangle_{X}+|V\rangle_{XX}|V\rangle_{X}$ and $\frac{1}{\sqrt2}(|early\rangle_{XX}|early\rangle_{X}+|late\rangle_{XX}|late\rangle_{X}$ for polarization and time-bin entanglement, respectively.}
\label{matrix}
\end{figure}

To quantify the nature and the degree of entanglement we performed several measurements. We firstly quantified the the entanglement in polarization without generating  the time-bin entanglement. Upon this we quantified the time-bin entanglement in the presence of polarization entanglement. To confirm the orthogonality of the two entangled degrees of freedom we performed a tomographic measurement of the complete hyper-entangled state. Finally, we quantified the polarization (time-bin) entanglement by performing a tomographic measurement averaged over all possible time-bin (polarization) projections.

The tomographic reconstruction of a bipartite state entangled either in time bin or polarization requires 16 projective measurements \cite{james}. To obtain projections in the polarization basis, we performed 16 measurements (the results of four such measurements are shown in Fig.~\ref{peaks}a), while for the analysis of time-bin entanglement we obtained the necessary 16 projections from 4 physical measurements (4 different phase settings of the analysis interferometers). This is possible because the $|\mathrm{early}\rangle$ and $|\mathrm{late}\rangle$ projection can always be clearly distinguished (time resolved) from the energy basis $|\mathrm{early\rangle+|late}\rangle$ ($|\mathrm{early\rangle-|late}\rangle$) \cite{takesue}. This is schematically depicted in Fig.~\ref{2p}d; the initial and the final peaks of photon arrival times reflect the classically correlated time basis while the middle of the three peaks represents the energy basis. In terms of coincidence events the measurement yields the five peaks shown in Fig.~\ref{peaks}b. The entanglement manifests itself as presence or absence of the middle peak as a function of the relative phase between the interferometers. The phase plate setting $\phi_{X(XX)}$ sets the projection for the energy basis to $|\mathrm{early\rangle+|late}\rangle$ for $0^\circ$ and to $|\mathrm{early\rangle-|late}\rangle$ for $90^\circ$ for each of the analysis interferometers separately.

For an independent measurement of the polarization entanglement we excluded the Michelson interferometers from the excitation and the detection. As result we obtained a concurrence value  $C_{p}=0.70(4)$. The corresponding fidelity to the maximally entangled state was $F_{p}=0.81(2)$. 
While for the polarization degree of freedom it was possible to perform an independent entanglement measurement this was not possible for the time-bin entanglement because because the quantum dot system used for this measurement does not offer the possibility to alter \cite{Trotta} the amount of fine structure splitting. Yet, we quantified the time-bin entanglement for one chosen polarization (HH projective measurement) and the results obtained yield the concurrence value of $C_{tb}=0.69(9)$. The corresponding fidelity was measured to be $F_{tb}=0.76(6)$.

The density matrix of a bipartite time-bin and polarization hyper-entangled state has dimension $16\times 16$ and therefore its estimation requires $256$ projective measurements. The result of such a measurement is shown in Fig. ~\ref{matrix}(a). The fidelity of the measured state to the state $|\Psi\rangle=(|\mathrm{H}\rangle_{\mathrm{XX}}|\mathrm{H}\rangle_{\mathrm{X}}+|\mathrm{V}\rangle_{\mathrm{XX}}|\mathrm{V}\rangle_{\mathrm{X}})\otimes(|\mathrm{early}\rangle_{\mathrm{XX}}|\mathrm{early}\rangle_{\mathrm{X}}+|\mathrm{late}\rangle_{\mathrm{XX}}|\mathrm{late}\rangle_{\mathrm{X}})$ yields $F_{hyp}=0.55(4)$. 

On the other hand, this hyper-entangled state is a product state of the polarization and the time-bin entangled state given and its subspaces can be reconstructed separately. %For our system this reconstruction still requires 64 projections, but by simultaneous measurement in two subspaces we can substantially reduce the measurement time while preserving the measurements statistics. %measurement time can be significantly shortened. 
We thus obtained the measurements in polarization basis by summing over the time basis projections and vice versa. This is equivalent to ignoring the polarization sub-system while measuring in time bin and vice versa. The real and imaginary part of the reconstructed density matrices for this measurement are plotted in Fig. ~\ref{matrix}(b). They yield the concurrence values $C_{p}=0.71(5)$ and $C_{tb}=0.76(8)$ of the polarization and time-bin entangled state, respectively, while the corresponding fidelities to the maximally entangled states were $F_{p}=0.81(6)$ and $F_{tb}=0.87(4)$.

{\em Conclusion}. Quantum networks will eventually need deterministic, single pair sources of entanglement, because in random sources, like spontaneous parametric down-conversion multiple-pair emissions scale with the emission rate and thus lead to an increasing error rate as the link length and thus the losses grow. Single quantum emitters, like quantum dots, on the other hand promise to deliver single entangled photon pairs. Our results show that it is possible to achieve high quality entanglement simultaneously in two degrees of freedom, ideally two ebits per emitted photon pair. This is not only a doubling of the resources per photon, but enables entirely new protocols and higher efficiency versions of others as discussed before. Already our source is not a random source and it remains to improve its error rate, which for the time-bin part is currently limited by the requirement to avoid multiple excitations. Using different state preparation schemes \cite{simon, hughes} that require an additional metastable state \cite{GershoniNature, GershoniScience} we can, in the future, turn this into a source of on-demand single hyper-entangled photon pairs.

\textbf{Methods}

The sample we used was grown by molecular beam epitaxy and consists of a layer of self-assembled In(Ga)As quantum dots in GaAs embedded within a distributed Bragg reflector $\lambda$-cavity. A bottom of the cavity consists of 15 pairs of AlAs/GaAs while the top has only a single pair. This cavity carries two functions: enhanced collection efficiency and reduced scattering of the excitation laser. During the measurements the sample was kept in a helium-flow cryostat temperature stabilized to 5 $\pm$~0.1~K. The excitation pulses were derived from an 82~MHz Ti:Sapphire laser.  The laser wavelength was 872.86~nm, which is half way between biexciton (873.57~nm) and exciton (872.17~nm) emission. To spectrally limit the scattered laser light and optimize the coherence of excitation, the pulse length was adjusted by means of a pulse-stretcher, which consisted of two diffraction gratings and a slit placed in-between them. After being coupled into a single mode fibre the laser light is sent through the unbalanced Michelson interferometer to generate the early and late pulses, which are sent towards the quantum dot sample via a single mode fibre. The excitation light was focused onto the sample from the side, while the emission was collected orthogonal to the excitation plane. The biexciton and exciton photons were spectrally separated in a home-built spectrometer and coupled into single mode fibres. The polarization analysis elements ($\lambda/4$ and $\lambda/2$ wave-plates and polarisers) were placed before the optical fibres. The fibre coupled biexciton and exciton emissions are sent into the interferometer for detection of time-bin entanglement. For the purposes of state tomography measurement, the relative phases between the interferometer for the biexciton and exciton photons and the the pump laser interferometer are controlled by phase plates. We quantified the degree of entanglement of the emitted photon pairs by measuring their quantum state through quantum state tomography \cite{james}. The lifetimes of the emitted photons were measured to be 220$\pm$20 and 400$\pm$20 ps for biexciton and exciton, respectively.

\textbf{Acknowledgments}

\begin{acknowledgments}
This work was funded by the European Research Council (project EnSeNa) and the Canadian Institute for Advanced Research through its Quantum Information Processing program. A. P. would like to thank the Austrian Science Fund for the support provided through project number V-375.  T. H. is receiving a DOC scholarship from the Austrian Academy of Sciences. P. M. would like to thank the Center for Integrated Quantum Science and Technology (IQST) for financial support.
\end{acknowledgments}

\end{document}